\documentclass[12pt]{article}
\usepackage{amsmath,amssymb}
\begin{document}

\centerline{\large A Simple Alternative to Jet-Clustering Algorithms}

\centerline{Howard Georgi\footnote{\noindent \tt georgi@physics.harvard.edu}}
\centerline{Center for the Fundamental Laws of Nature}
\centerline{Jefferson Physical Laboratory}
\centerline{Harvard University, Cambridge, MA 02138}

\date{\today}

\begin{abstract}
I describe a class of iterative jet algorithms that are
based on maximizing a fixed function of the total 4-momentum 
rather than clustering of pairs of jets. I describe some of the properties
of the simplest examples of this class, appropriate for jets at an $e^+e^-$
machine.  These examples are
sufficiently simple that many features of the jets that they define can be
determined analytically with ease.   The jets constructed in this way have
some potentially useful properties, including a strong form of infrared safety.
\end{abstract}



Iterative jet clustering algorithms have become an important tool in the
analysis of high-energy scattering experiments (see for example
\cite{Dasgupta:2013ihk}, and references therein).  
In this note, I describe a class of iterative jet algorithms that are
based on maximizing a fixed function of the total 4-momentum 
rather than clustering of pairs of jets. I describe some of the properties
of the simplest examples of this class.  These examples are
sufficiently simple that many features of the jets they define can be
determined analytically with ease.   The jets constructed in this way have
some potentially useful properties, including a strong form of infrared safety.

The idea can be stated very simply.
Suppose that we have a collection of 4-momenta
$p^\mu_j$ that we want to organize into jets.  In practice, we will
typically be interested in 
masses $\sqrt{p_j^\mu p_{j\mu}}$ that are small compared to their energies and can
be ignored in leading order.  
This is not necessary for the construction, but it leads to considerable
simplification, and we will assume that we can set all the particle masses
to zero.  The jets will then be
particular sets of momenta, $\alpha$ with total jet momenta
\begin{equation}
P^\mu_\alpha=\sum_{j\in\alpha}p_j^\mu
\end{equation}
The underlying assumption is that we want to combine a collection of lines into a
single jet, hence increasing the jet energy, if doing so does not increase
the jet mass to much.  So we choose a ``jet function''  $J(P_\alpha)$ of the total
4-momentum of the ensemble that depends
only on the total energy, $P_\alpha^0$ and the total mass squared
divided by energy:
\begin{equation}
J(P)=f(P^0,P^\mu P_\mu/P^0)
\label{jetfunction}
\end{equation}
We require $J(P)$ to have the
property that it
increases with increasing energy and decreases with increasing $m^2/E$.
We then
{\bf find the set $\alpha$ with the maximum value of $J$.} This gives us our
highest $J$ jet (note that it is not necessarily the highest energy jet).  Then, as
usual in an iterative jet construction, the 
lines in $\alpha$ are removed and the process is repeated until no lines
are left.  
What makes this different from any pair-wise clustering algorithm that I
know of is that we maximize over all possible clusters all at once, rather
than building up the jet by clustering pairs.  One might worry that this
will make the procedure unwieldy for events with many particles.  But we
will see that at least for one very simple form of the jet function, the
clustering is local so that the algorithm can be implemented efficiently.
Furthermore the boundaries between the jets have very simple properties
that I believe will lead to important simplifications in perturbative
calculations (and perhaps beyond).\cite{Kelley:2012zs}

I should emphasize that in this note, with a jet function like
(\ref{jetfunction}) that depends on energy and mass, rather than transverse mass,
I am illustrating the idea for jets at an $e^+e^-$ machine,
ignoring (for simplicity and because I am not sure how to handle it) the
additional complication of hadron beams.
I will discuss what I believe is the simplest example of this
scheme, in which the function has the form
\begin{equation}
J_\beta(P)=P^0-\beta P^\mu P_\mu/P^0
\;\;\mbox{for}\;\;\beta>1
\end{equation}
This is monotonically increasing in $E$ and decreasing in
$m^2$.\footnote{This is also true for $0<\beta\leq1$, but we need $\beta>1$
for the analysis below.}  As we will see, this produces jets with 
no fixed ``cone size'', but as $\beta$ increases, there is
more of a penalty for large jet 
mass, and so the jets become more collimated and
effectively there is a cone size that decreases as $\beta$ increases.
Obviously, for $\beta=0$, 
everything gets included in one ``jet'', so this is not particularly
interesting.  But $\beta>1$ is interesting
and we will be able to understand why analytically. 

Let us first
consider some general properties of the jet with the largest
$J_\beta$.  Iterating this will give us interesting information about all
the jets.

So we suppose that 
\begin{equation}
P_\alpha^\mu\equiv \sum_{j\in\alpha}p_j
\end{equation}
maximizes $J_\beta$.  
It is obvious that
\begin{equation}
\beta\,P_\alpha^2<{P_\alpha^0}^2
\label{p2less}
\end{equation}
We must have $J_\beta(P)>0$, because there are
always positive $J_\beta$s (for example single particles) and
(\ref{p2less}) follows immediately.
Then we can write
\begin{equation}
(\beta-1){P_\alpha^0}^2<\beta\,P_\alpha^2
\end{equation}
where $P_\alpha=|\vec P_\alpha|$.
Hence
\begin{equation}
P_\alpha>\sqrt{\frac{\beta-1}{\beta}}
\,P_\alpha^0
\label{nearlylightlike}
\end{equation}
This shows that for large $\beta$, the jets are necessarily nearly light-like.   

If there is only one line in the jet, (\ref{nearlylightlike}) is
automatically satisfied for any $\beta>1$.
Suppose that there
is more than one line in $\alpha$ and consider a line with 4-momentum 
$p_j^\mu$  for any
$j\in\alpha$.  Because both 
$p_j^\mu$ and the rest of the 4-momentum of the jet, $P_\alpha^\mu-p_j^\mu$
have lower $J_\beta$ than
$P_\alpha^\mu$ by assumption, we can write
\begin{equation}
J_\beta(P_\alpha)>\max\left(J_\beta(P_\alpha-p_j)\,,\,J_\beta(p_j)\right)
\label{bothgtr}
\end{equation}
Let $z$ be the cosine of the angle between
$\vec p_j$ and the jet direction $\vec P_\alpha$.
It is also convenient to define a ``jet velocity''
\begin{equation}
v_\alpha\equiv P_\alpha/P_\alpha^0
\end{equation}
and the fraction of the jet energy carried by line $j$
\begin{equation}
r_j\equiv E_j/P_\alpha^0
\end{equation}
From (\ref{nearlylightlike}), we know
\begin{equation}
\sqrt{\frac{\beta-1}{\beta}}<v_\alpha<1
\end{equation}
and of course from energy conservation, 
\begin{equation}
0<r_j<1
\end{equation}
In terms of $v_\alpha$ and $r_j$, (\ref{bothgtr}) becomes
\begin{equation}
1-\beta\,(1-v_\alpha^2)
>\max\left(
1-r_j-\beta\frac{1-v_\alpha^2-2r_j(1-z\,v_\alpha)}{1-r_j},r_j\right)
\label{bothgtr2}
\end{equation}
This immediately implies a stronger bound on $v_\alpha$ that
(\ref{nearlylightlike}):
\begin{equation}
P_\alpha/P_\alpha^0=v_\alpha>\sqrt{1-\frac{1-r_j}{\beta}}
\label{morelightlike}
\end{equation}
This is not a surprise.  It is obvious that if the jet contains a massless
particle carrying most of the energy, it is nearly light-like.  It is also
clear that as $r_j\to1$, $z\to1$, just by 4-momentum conservation.  This
means that the bound on jet ``size'' in the sense of the largest possible angle of a
particle in the jet from the jet direction is determined by the soft
particles in the jet.

(\ref{bothgtr2})  also gives
\begin{equation}
z>\frac{\beta(1+v_\alpha^2)-(1-r_j)}{2\beta v_\alpha}
\label{maxz}
\end{equation}
or in terms of the angle $\theta$ between $\vec p_j$ and the jet direction
\begin{equation}
2\sin\frac{\theta}{2}<\sqrt{\frac{1-r_j-\beta(1-v_\alpha)^2}{\beta v_\alpha}}
\label{mintheta}
\end{equation}
So the maximum angular size of the jet is obtained for soft lines,
$r_j\to0$, and also 
depends on the jet velocity.  The maximum angular size of the jet is given by 
\begin{equation}
2\arcsin\left(\frac{1}{2\sqrt{\beta}}\right)
\;
\mbox{as}\; v_\alpha\to1
\label{zmin1}
\end{equation}
and it goes to
\begin{equation}
2\arcsin\Bigl(1-\sqrt{1-1/\beta}\Bigr)
\;
\mbox{as}\; v_\alpha\to\sqrt{1-1/\beta}
\label{zmin}
\end{equation}

Summarizing the most important result so far, we have found that 
all the lines in the highest $J_\beta$ jet are inside a cone of angle
\begin{gather}
\Theta(\beta,v_\alpha)\equiv 
2\arcsin\left(
\sqrt{\frac{1-\beta(1-v_\alpha)^2}{4\beta v_\alpha}}\right)
\label{theta}
\\
\leq2\arcsin\Bigl(1-\sqrt{1-1/\beta}\Bigr)
\label{theta1}
\end{gather}
around the jet direction.  

We now
go on to discuss the particles that are NOT in the highest $J_\beta$ jet.

For a particle with $p_j^\mu$ for $j\not\in\alpha$, the relations look very
similar to (\ref{bothgtr}, \ref{bothgtr2}). 
\begin{equation}
J_\beta(P_\alpha)>\max\left(J_\beta(P_\alpha+p_j)\,,\,J_\beta(p_j)\right)
\label{obothgtr}
\end{equation}
or
\begin{equation}
1-\beta\,(1-v_\alpha^2)
>\max\left(
1+r_j-\beta\frac{1-v_\alpha^2+2r_j(1-z\,v_\alpha)}{1+r_j},r_j\right)
\label{obothgtr2}
\end{equation}
Note that in this case, while energy conservation does not require $r_j<1$,
(\ref{obothgtr2}) does.

(\ref{bothgtr2})  also gives
\begin{equation}
z<\frac{\beta(1+v_\alpha^2)-(1+r_j)}{2\beta v_\alpha}
\label{minz}
\end{equation}
or in terms of the angle $\theta$ between $\vec p_j$ and the jet direction
\begin{equation}
2\sin\frac{\theta}{2}>\sqrt{\frac{1+r_j-\beta(1-v_\alpha)^2}{\beta
    v_\alpha}}
>2\sin\frac{\Theta(\beta,v_\alpha)}{2}
\label{maxtheta}
\end{equation}
This immediately implies that all other particle lines are outside the cone
of angular size $\Theta(\beta,v_\alpha)$
around the highest $\beta$ jet.
Furthermore, the particles not in the jet can only approach the
jet boundary as $r_j\to0$ --- that is only for
infinitely soft particles.

Thus even though we did not impose a cone size, the
jet function $J_\beta$ did so for us, at least for the highest $J_\beta$
jet.  Particles in the jet are inside the cone.  Particles not in the jet
are outside the cone.
The cone size $\Theta(\beta,v_\alpha)$ varies slightly over the allowed range of
$v_\alpha$, but for large $\beta$ goes to $1/\sqrt{\beta}$.

Because of this clean separation between the jet defined by $J_\beta$ and
all the other particles,
iterating the procedure is absolutely no problem.  The jets produced by the
procedure are all non-overlapping and bounded by cones of size less than
$\Theta(\beta,v_\alpha)$ (each with its own $v_\alpha$ of course).   
This also insures the IR safety of this procedure.  All particles in a
particular direction must obviously be in the same jet.  In fact, the
procedure is VERY IR safe, because particles near the boundary of the cone must
get arbitrarily soft.

The relation (\ref{zmin}) means that the jets are localized for large
$\beta$, so one does not have to calculate $J_\beta$ for all subsets.
Instead, the $4\pi$ solid angle can be divided into fiducial regions in any
convenient way, and the jets in each fiducial region can be found by
looking only at the fiducial region plus the border region within an angle
$2\arcsin\Bigl(1-\sqrt{1-1/\beta}\Bigr)$ of the boundary.  This
procedure will also give fake jets with direction outside the fiducial
region, which should simply be ignored. But for large $\beta$, this should
enable an enormous improvement in jet-finding efficiency.

I hope I have convinced the reader that the new jet algorithm I am
proposing is simple and interesting.  It remains to be seen whether it is
useful.   In closing it is worth emphasizing again that
since we have considered a jet function that depends on energy and mass, we
are focusing here on a situation like that at an $e^+e^-$ machine where all
the jets are on the same footing.  Hadron beams and their
associated jets introduce an additional complication.  It is important to
try and generalize this method to hadron colliders because, at least
for the moment, that is where all the action is.   

\subsubsection*{Acknowledgements}
I am very grateful to Matthew Schwartz and Ilye Feige for many
informative conversations about jets, to Steve Ellis for a reality
check and to Jakub Scholtz for catching a minor misstatement in the first
version, and to Jesse
Thaler for helpful comments.  I am 
also extremely grateful to the  
{\bf Invisibles}, including Belen Gavela and Pilar Hernandez whose
invitation to the Invisibles Summer School and Workshop was an important
motivation for this work.  This work grew initially out of a three-year-long
effort with Greg Kestin and Aqil Sajjad to construct an effective field
theory on the light shell.  I am grateful for their help during this
ongoing process.
This research has been supported in part by the National Science Foundation
under grants PHY-1067976 and PHY-1418114. 

\bibliography{jets}

\end{document}